\documentclass[journal=jacsat,manuscript=article]{achemso}
\usepackage[version=3]{mhchem} % Formula subscripts using \ce{}
\usepackage[T1]{fontenc}       % Use modern font encodings
\usepackage{amsmath}
\usepackage{color}

\newcommand{\UNSW}
	{School of Electrical Engineering and Telecommunications, The University of New South Wales, Sydney NSW 2052 Australia}
\newcommand{\Purdue}
	{Department of Electrical and Computer Engineering, Purdue University, West Lafayette, IN 47907, USA}	

\author{Kok Wai Chan}\affiliation[University of New South Wales]{\UNSW}
\author{Harshad Sahasrabudhe}\affiliation[Purdue University]{\Purdue}

\author{Wister Huang}\affiliation[University of New South Wales]{\UNSW}

\author{Yu Wang}\affiliation[Purdue University]{\Purdue}

\author{Henry C. Yang}
\author{Menno Veldhorst}
\altaffiliation{Current address: QuTech and Kavli Institute of Nanoscience, Delft University of Technology, PO Box 5046, 2600 GA Delft, The Netherlands}
\author{Jason C. C. Hwang}
\author{Fahd A. Mohiyaddin}
\altaffiliation{Current address: IMEC, Kapeldreef 75, Leuven 3001, Belgium}
\author{Fay E. Hudson}
\affiliation[University of New South Wales]{\UNSW}

\author{Kohei M. Itoh}
\affiliation[Keio University]
{School of Fundamental Science and Technology, Keio University, 3-14-1 Hiyoshi, Kohoku-ku, Yokohama 223-8522, Japan}

\author{Andre Saraiva}
\author{Andrea Morello}
\author{Arne Laucht}
\affiliation{\UNSW}

\author{Rajib Rahman}
\affiliation{\Purdue}
\affiliation[physics]{School of Physics, The University of New South Wales, Sydney, NSW 2052, Australia}
\email{rajib.rahman@unsw.edu.au}

\author{Andrew S. Dzurak}
\email{a.dzurak@unsw.edu.au}
\affiliation{\UNSW}

\title[An \textsf{achemso} demo]{Exchange coupling in a linear chain of three quantum-dot spin qubits in silicon}

\abbreviations{}

\begin{document}
\begin{abstract}
Quantum gates between spin qubits can be implemented leveraging the natural Heisenberg exchange interaction between two electrons in contact with each other. This interaction is controllable by electrically tailoring the overlap between electronic wavefunctions in quantum dot systems, as long as they occupy neighbouring dots. An alternative route is the exploration of superexchange -- the coupling between remote spins mediated by a third idle electron that bridges the distance between quantum dots. We experimentally demonstrate direct exchange coupling and provide evidence for second neighbour mediated superexchange in a linear array of three single-electron spin qubits in silicon, inferred from the electron spin resonance frequency spectra. We confirm theoretically through atomistic modeling that the device geometry only allows for sizeable direct exchange coupling for neighbouring dots, while next nearest neighbour coupling cannot stem from the vanishingly small tail of the electronic wavefunction of the remote dots, and is only possible if mediated.
\end{abstract}

Spin qubits in silicon are gaining increasing attention due to recent demonstrations of high fidelity control~\cite{yoneda2018quantum,yang2019silicon,xue2019benchmarking,zajac2018resonantly,huang2019fidelity}. Extremely long coherence times have been achieved in isotopically purified silicon~\cite{veldhorst2014addressable,muhonen2014storing,eng2015isotopically,tyryshkin2012electron}, such that qubits in quantum dots can be constructed with fidelities nearing fault-tolerant thresholds~\cite{helsen2018quantum}. Electrostatically defined quantum dots enable tunable control of the electronic wavefunction, so that qubits can be deterministically formed and coupled. While silicon has small intrinsic spin-orbit coupling, the presence of a sharp interface together with the extremely narrow electron spin resonance (ESR) linewidth in enriched $^{28}$Si, enables $g$-factor tuning for individual qubit control~\cite{veldhorst2014addressable,huang2018fidelity}. These advantages have contributed to the significant progress over the past few years in the demonstration of silicon qubit systems~\cite{veldhorst2014addressable,veldhorst2015two, yoneda2018quantum,zajac2018resonantly,watson2018programmable,huang2019fidelity,yang2019silicon,petit2019universal}. 

Looking forward to a larger processing unit, the issue of qubit connectivity arises~\cite{vandersypen2017interfacing}. While exchange is a natural resource for the entanglement of spin qubits, it relies on the Pauli exclusion principle, which means it is only large enough if the electrons are in direct contact with each other. Since the tunnel coupling between quantum dots decays exponentially with their separation, exchange-controlled two-qubit gates have only been executed between neighbouring quantum dots~\cite{veldhorst2015two,zajac2018resonantly,watson2018programmable,huang2019fidelity}. While local single and two-qubit control provides a universal gate set for quantum computation, long-range links increasing the qubit connectivity are highly advantageous for quantum error correction as well as for reducing the number of operations required in quantum algorithms, by reducing the need for intermediate swap operations. While such long-range links may not be feasible using the direct exchange interaction, two remote spins may have significant interaction due to virtual exchange through a central mediator. This mechanism is called superexchange and is present in various magnetic materials. Early theoretical studies confirm that a significant reduction in the number of quantum operations is obtained if this mediated exchange coupling is harnessed for coupling distant qubits~\cite{fei2012mediated, deng2020interplay}.

Here, we realize a three-qubit system using silicon quantum dots~\cite{veldhorst2014addressable,veldhorst2015two} fabricated in a linear array and study the exchange couplings between them. In addition to measuring the direct exchange between adjacent qubits $J_{\rm direct}$, we are able to extract a non-vanishing  superexchange coupling $J_{\rm super}$ between next-nearest neighbour qubits~\cite{baart2017coherent,malinowski2018fast}. We further study the dependence of direct exchange coupling $J_{\rm direct}$ on the critical dimensions of the structure with the aid of a tight binding atomistic simulation tool, NEMO~\cite{klimeck2002computer,sahasrabudhe2018optimization,tankasala2018two}. The simulations are geared towards calculating the two-electron ground state in the dots by modeling the electrostatic landscape, material properties and exchange-correlation effects in an optimized parametric space. The parameters required as inputs are derived from experiments, such as device geometry, gate voltages, and impurity concentrations, and from ab-initio methods, e.g. bandstructure of silicon and band gap offsets. Our findings confirm the validity of the interpretation of a direct exchange among neighbouring dots. The computational cost of describing strongly correlated multielectron states grows exponentially with the number of electrons, which hindered a comparison between simulations and experiments in the case of mediated exchange.

\begin{figure}
	\includegraphics[width=7.46cm]{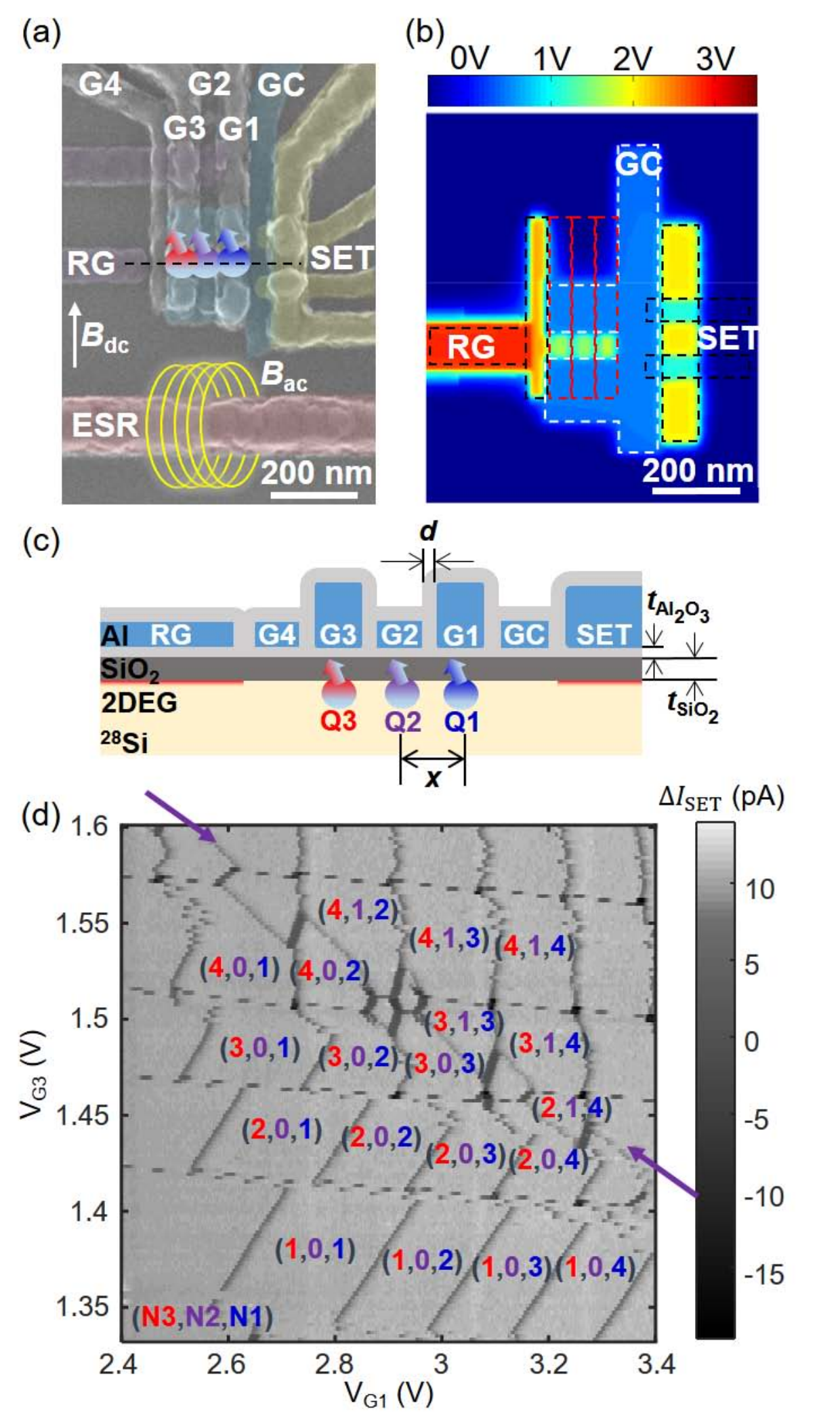}
	\caption{SEM image, Finite element modelling (FEM) of electrostatics, and charge stability diagram of the silicon three-qubit device.
		(a) SEM of a fabricated device. (b) FEM simulation of the device's electrostatic potential profile. This profile is used in the exchange coupling simulation with NEMO. The device consists of qubits Q1, Q2, and Q3 underneath the gates G1, G2, and G3, respectively. Electrons in the quantum dots are loaded from the reservoir gate RG, with G4 acting as tunnel barrier and GC as confinement gate at low voltages. SET is the single-electron charge sensor.
		(c) Cross sectional schematic along the black dashed line in (a) showing the active region of the device, where the 2DEG is formed and quantum dots are confined. Here, $t_{\ce{SiO2}}$ is the thickness of silicon dioxide, $t_{\ce{Al2O3}}$ is the thickness of aluminum oxide on top of the gate oxide, $d$ is the thickness of aluminum oxide in between the qubit gates, and $x$ is the mean centre-to-centre distance between qubits used for exchange coupling simulations. (d) Charge stability diagram of the triple quantum dot system as a function of voltages applied to the gates G3 and G1. The electron occupancy on each qubit is assigned as (N3,N2,N1) as depicted in the diagram. The diagonal transition line, marked with purple arrows, is the loading of qubit Q2 with one electron. The quantum dot Q2 has zero (one) electron occupancy below (above) of this transition.
	}
	\label{fig1}
\end{figure}

Figures~\ref{fig1}(a) and (b) show the fabricated and modelled three-qubit device, respectively. The device is fabricated on an isotopically enriched $^{28}$Si epilayer, with a residual $^{29}$Si concentration of 800 p.p.m.~\cite{Kohei28SiMRS2014}, in which dephasing effects due to the nuclear spin bath are greatly suppressed, increasing the coherence time of the electron spins~\cite{veldhorst2014addressable}. Figure~\ref{fig1}(a) is the scanning electron micrograph (SEM) of a nominally identical device at the active region where the qubits and their readout/control circuitry are defined. The aluminum gates are highlighted and labeled in the SEM image. The qubit gates G1, G2 and G3 control the electron occupancies of the three quantum dots. A single electron transistor (SET) serves as charge sensor with two tunnel barrier gates underneath a top gate~\cite{AngusNano07,morello2010single}. Tunneling from the SET to the dot under G1 is prevented by the confinement gate GC. Electron spin resonance (ESR) experiments are performed with an on-chip microwave transmission line. In Figure~\ref{fig1}(b), we show the results from a FEM simulation of the electrostatic potential profile. This was used as an input to the atomistic full configuration interaction method in NEMO. Figure~\ref{fig1}(c) is a cross sectional schematic along the black dashed line drawn in Figure~\ref{fig1}(a). In this schematic, we illustrate the position of qubits Q1, Q2 and Q3 underneath gates G1, G2 and G3. Gate G4 acts as a tunnel barrier for loading electrons from the two-dimensional electron gas (2DEG) formed below gate RG. Note that the 2DEG reservoir (RG) and SET have independent ohmic contacts. In Figure~\ref{fig1}(c), we also label the parameters used in NEMO to simulate $J_{\rm direct}$ between the qubits. These are the thickness of silicon dioxide ($t_{\ce{SiO2}}$), thickness of aluminum oxide on top of the gate oxide ($t_{\ce{Al2O3}}$), thickness of aluminum oxide in between the G gates ($d$) and distance between the centres of the quantum dots ($x$).

The device was measured in a dilution refrigerator at a base temperature of 40~mK and dc magnetic field, $B_{\rm dc}$ = 1.4~T. The charge stability diagram of the triple quantum dot system as a function of G3 and G1 gate voltages is shown in Figure~\ref{fig1}(d). The electron occupancy of qubits (Q3,Q2,Q1) is labelled as (N3,N2,N1) in the stability diagram. As $V_{\rm G1}$ ($V_{\rm G3}$) increases in voltage and passes the charge transitions, the electron occupancy in Q1 (Q3) increases. The diagonal transition marked by purple arrows indicate the charge transition for Q2. Q2 has zero electron occupancy below and one-electron occupancy above of this transition. In this device, we are able to perform ESR on all three qubits in the lower valley state (1 electron loaded), and additionally in the upper valley state for Q1 and Q3 (3 electrons loaded)~\cite{yang2019silicon}. Their Ramsey dephasing times, $T_{2}^*$ are shown in the supplementary information. The qubits are highly tunable with electric and magnetic fields~\cite{Veldhorst_SOC_PRB2014}. Detailed analysis and theoretical studies of their Stark shift dependencies on d.c.~magnetic field direction with respect to the silicon crystal orientation are available in Ref.~\cite{Rifat}.  

%We have compared one-electron and three-electron qubits and show non-trivial Stark shift tuning, possibly due to steps at the oxide interface~\cite{WisterPRB2017,Veldhorst_SOC_PRB2014}

%Message 2

%By lifting the Q1 electrochemical potential towards the Q2-Q1 anti-crossing through electrostatically controlled gate G1, the resonance frequency of Q2 diverts as a spin funnel as shown in Figure~\ref{fig2}(a). 

We now demonstrate the ability to couple the qubits. In addition to the exchange coupling between Q2-Q1 at the (1,1)--(0,2) transition (Figure~\ref{fig2}(a)), previously discussed in ref.~\citenum{veldhorst2015two}, we have coupled Q3-Q1 at the (3,1,3)--(2,1,4) transition (Figure~\ref{fig2}(c)) with Q2 acting as a mediator. All experiments are performed with a similar protocol, where a voltage pulse $V_{\rm P}$, applied to G1, controls the detuning energy $\epsilon$ between the quantum dots to bring the Q1 electrochemical potential near the (1,1)--(0,2) anti-crossing (or equivalent). 

In Figure~\ref{fig2}(a), we initialize Q2-Q1 with opposite spins \newcommand {\ket} [1] {|{ #1 \rangle}}$\ket{\uparrow\downarrow}$ or $\ket{\downarrow\uparrow}$, followed by a microwave pulse and spin readout on Q2. This generates a characteristic plot that maps the resonance shifts caused by exchange coupling, called spin funnels. We describe next the details of initialisation, spin resonance and readout for this experiment.

Figure~\ref{fig2}(b) is the G2-G1 double-dot charge stability diagram marking the positions of microwave control (C2) and spin readout (R2) of Q2. The two qubit dots are schematically represented. In this experiment, the voltage at gate G4 is set to be positive enough such that the reservoir extends under it and the dot Q2 is separated from the 2DEG using G3 as a barrier gate. The state initialization into anti-parallel spins is done by pulsing the system deep into the (1,2) electron configuration for 100$\mu$s, marked as I$_{\rm s}$, so that Q1 is loaded with 2 electrons in a singlet state. Then, the system is rapidly brought back into the (1,1) region where one electron from Q1 is unloaded into the reservoir through the (0,2) intermediate state. We note that the plot shows exclusively spin transitions occurring in Q2 -- the difference in resonance frequencies between Q1 and Q2 is approximately 40 MHz, much larger than the frequency shifts plotted in Fig.~\ref{fig2}(a). The downward (i) and upward (ii) bending spin funnels map out the energy difference between the different spin states as shown in Figure~\ref{fig2}(e). The experimental mapping of the resonance frequencies was performed up to the point where dephasing due to charge noise became too strong. The tunnel coupling, $t_0$ between Q2-Q1 is estimated to be 900~MHz through fitting, with a linear gate dependence of the resonance frequency (Stark shift) of 19~MHz/V on both sides of the spin funnel~\cite{Veldhorst_SOC_PRB2014,veldhorst2015two}. The controlled-NOT (CNOT) quantum gate operation with these qubits has been reported in ref.~\citenum{veldhorst2015two}.

\begin{figure}
	\includegraphics[width=17cm]{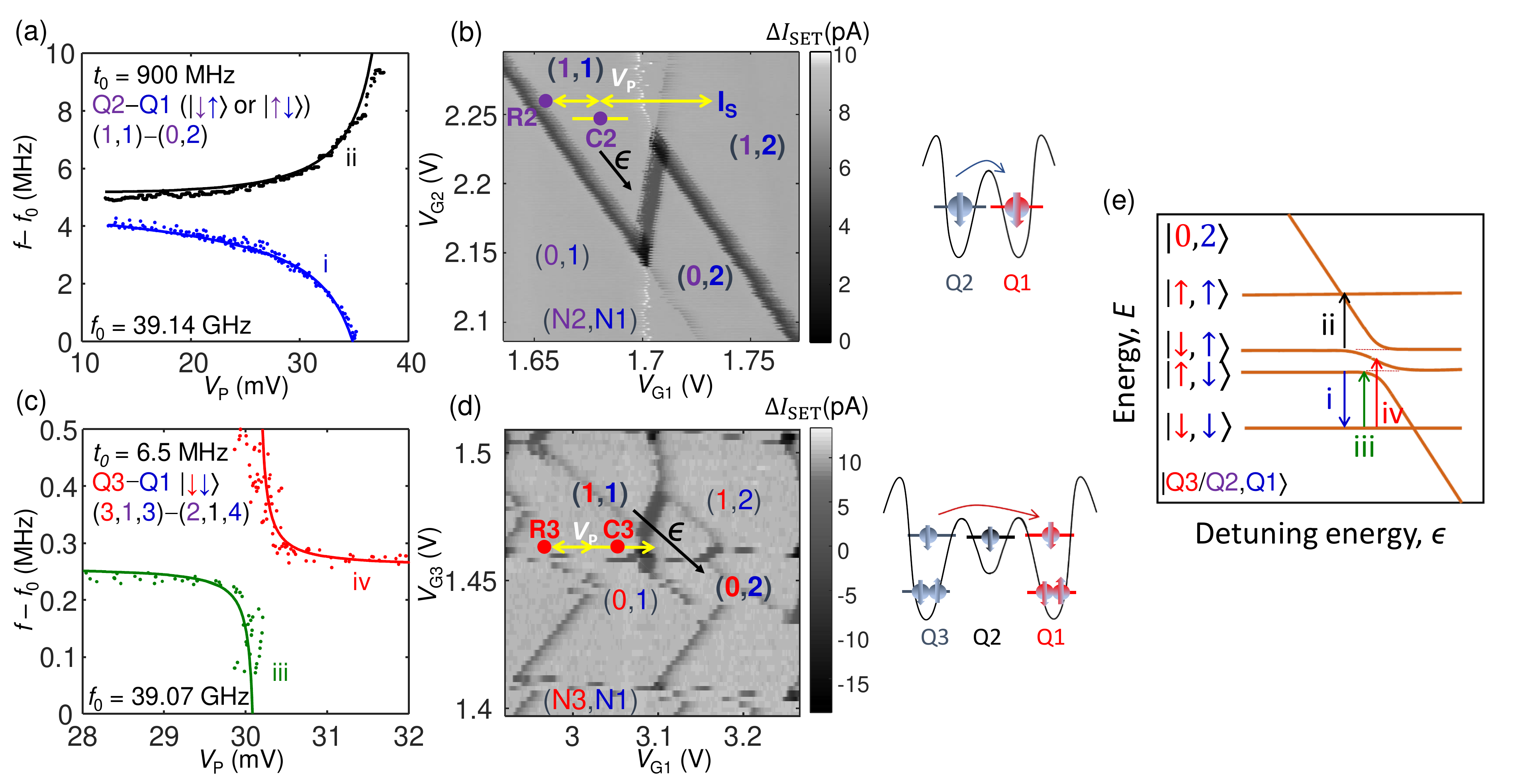}
	%  As well as the standard float types \texttt{table}\\
	%  and \texttt{figure}, the class also recognises\\
	%  \texttt{scheme}, \texttt{chart} and \texttt{graph}.
	\caption{Experimental nearest neighbour and next-nearest neighbour couplings in a three-qubit system.
		Measured ESR funnels on (a) Q2 due to exchange coupling and (c) Q3 due to possible super-exchange with Q1. Tunnel coupling for Q2-Q1 and effective tunnel coupling for Q3-Q1 are fitted to 900~MHz and 6.5~MHz, respectively. (a) and (c) have been offset by $f_0$ = 39.14~GHz and 39.07~GHz. For clarity, only the minima and maxima spin up probabilities of each frequency trace are plotted to elucidate the spin-down and spin-up funnels. (b) Charge stability diagram showing the regime where Q2-Q1 are exchange coupled, such as represented schematically to the right of the plot. (d) Detailed charge stability diagram from Figure~\ref{fig1}(d) near the region where we have studied the super-exchange coupling of Q3-Q1, such as represented schematically to the right of the plot. Here, we marked the Q3 spin readout (R3) and ESR control (C3) positions on the stability diagram. (e) Schematic energy level diagram for a $\ket{1,1}$ charge state anticrossing with a $\ket{0,2}$ charge state as a function of detuning energy, $\epsilon$. The roman numbers point out the energy difference between the spin states that are mapped on the resonance frequency spectrum in (a) and (c).
	}
	\label{fig2}
\end{figure}

Next, Figure~\ref{fig2}(c) shows the spin funnels for Q3 when coupled to Q1, at the (3,1,3)--(2,1,4) transition with Q2 acting as mediator. In this electron configuration the coupled qubits are occupying the upper valley states, where the first two electrons form a closed shell within the lower valley state in each dot. To simplify the analysis, we omit Q2 and the first two electrons which are loaded into the lower valley state of both Q3 and Q1. As such, we end up with a (1,1)--(0,2) transition for the Q3-Q1 interaction in the upper valley, similar to the Q2-Q1 coupled system in the lower valley. This is represented in Fig.~\ref{fig2}(d). Here, we initialize the states of Q3 and Q1 to both spin down, $\ket{\downarrow\downarrow}$. We note that this initialization is performed by setting the Q3 dot potential near the (3,1,3)--(2,1,3) transition in a way that the reservoir chemical potential lies between the Zeeman split spin states and waiting 100 $\mu$s, which leads to an efficient flushing of any residual spin up state and loading of a spin down electron in Q3. The spin in Q1 is initialized by relaxation to the spin down ground state and is not excited during the experiment.

Again, by applying microwave pulses, we obtained the resonance frequency by observing the spin up fraction of Q3 when the spin state of Q1 stays down. The Q3 spin readout and initialization (R3) and ESR control (C3) positions are shown in Figure~\ref{fig2}(d). The green downward (iii) bending funnel maps to the energy difference between $\ket{\downarrow\downarrow}$ and $\ket{\uparrow\downarrow}$ states in Figure~\ref{fig2}(e) and as we pulse over the anti-crossing, we map the red spin funnel coming from the top (iv) due to a superposition of $\ket{\uparrow\downarrow}$ and $\ket{\downarrow\uparrow}$. The effective tunnel coupling between Q3 and Q1 is fitted to 6.5~MHz using the same Stark shift as before. As expected, the indirect coupling through the mediating quantum dot is much lower than the direct nearest neighbour coupling.

%Message 3

\begin{figure}
	\includegraphics[width=8.46cm]{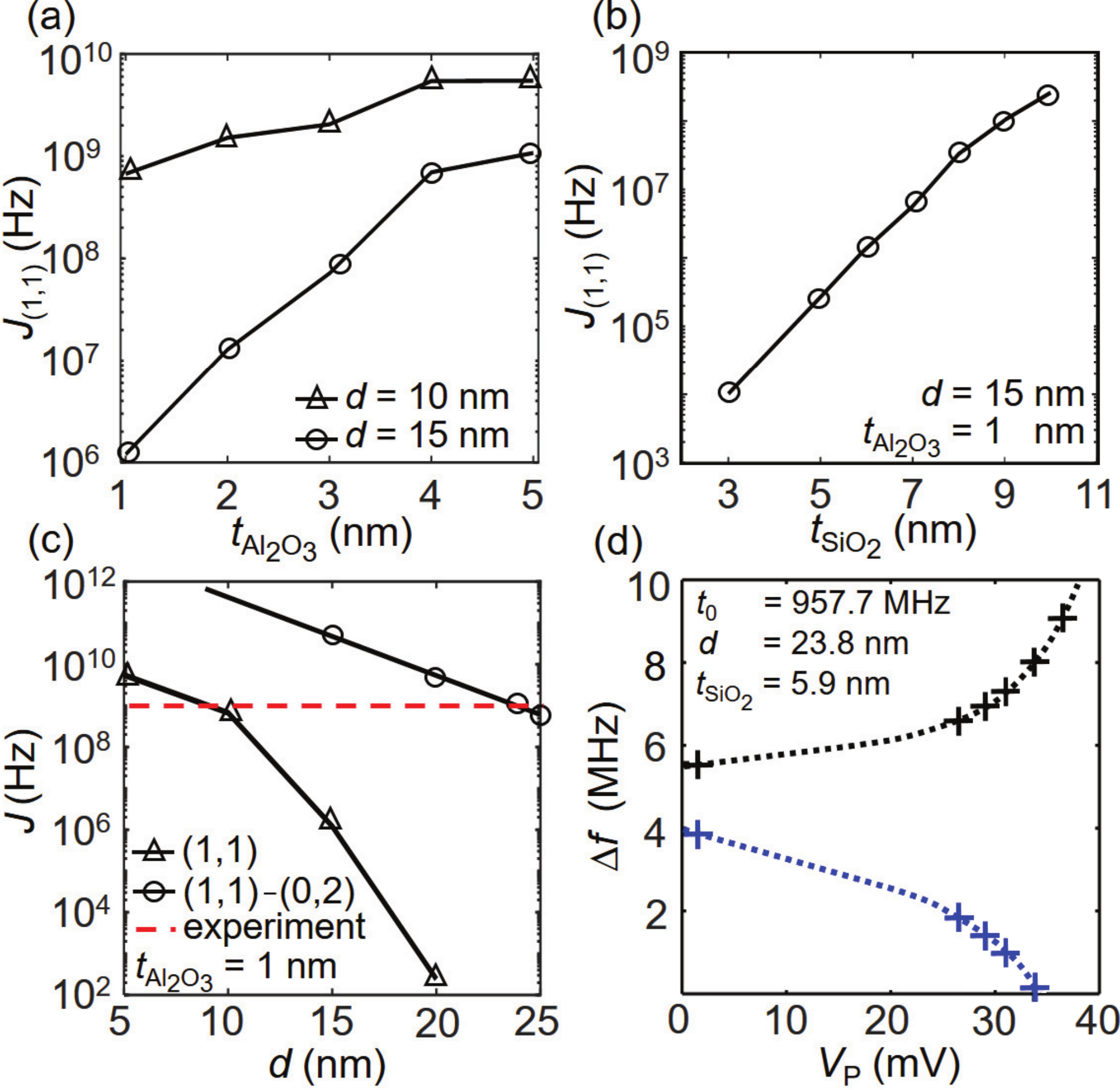}
	%  As well as the standard float types \texttt{table}\\
	%  and \texttt{figure}, the class also recognises\\
	%  \texttt{scheme}, \texttt{chart} and \texttt{graph}.
	\caption{NEMO simulation of $J_{\rm direct}$ coupling between Q2-Q1.
		Simulated $J_{\rm direct}$ coupling in the (1,1) electron configuration as a function of (a) $t_{\ce{Al2O3}}$ and (b) $t_{\ce{SiO2}}$.
		(c) Simulated $J_{\rm direct}$ coupling in the symmetric double dot (1,1) configuration (triangles) and in the detuned configuration, near the  (1,1)--(0,2) transition, as a function of $d$. As expected, we observe that the coupling between qubits decreases as $d$ increases. In order to match the experimental fitted tunnel coupling of 900~MHz, marked by the red dashed line, we picked $d$ = 23.8~nm.
		(d) Simulated spin funnels to match the experimental measured results shown in Figure~\ref{fig2}(a). The simulated spin funnels are obtained with $x$ = 36~nm, $t_{\ce{SiO2}}$ = 5.9~nm, $t_{\ce{Al2O3}}$ = 1~nm and $d$ = 23.8~nm. Calculated tunnel coupling here is 957.7~MHz, very close to the experimental $t_0$ = 900~MHz. 
	}
		%We notice that $d$ is significantly higher here and very unlikely as the aluminum oxide between gates should not grow beyond 10~nm. This indicates that there could be strain induced on the oxide interface caused by aluminum gates deposited at high temperature but measurement done at mK. Such strain lifted the tunnel barrier and can be represented with larger $d$ in the simulation.
		
	\label{fig3}
\end{figure}

As part of the effort to scale up this qubit prototype, further analysis on the exchange coupling between two nearest neighbour qubits is performed using NEMO to aid in the design of future device architectures. This analysis entails an atomistic model of the quantum dot device, with electrostatic potential from the gates obtained from FEM simulations. The electron-electron interactions, including exchange and correlation, are taken into account using a configuration interaction approach. We simulate the effect of varying the device specifications on $J_{\rm direct}$ between Q2 and Q1. The main parameters varied in our study are the oxide dimensions $t_{\ce{SiO2}}$,  $t_{\ce{Al2O3}}$ and $d$ which are described in Figure~\ref{fig1}(c). These parameters are set by the fabrication process and may be designed differently from device to device to optimize qubit performance. In Figure~\ref{fig3}(a), we plot $J_{\rm direct}$ in the symmetric (1,1) electron configuration as a function of $t_{\ce{Al2O3}}$ for a gate separation of $d$ = 10~nm (triangles) and 15~nm (circles). The silicon dioxide thickness $t_{\ce{SiO2}}$ is kept constant at 5.9~nm, consistent with such devices. We highlight that the gate separation $d$ is changed keeping the distance between dot centres $x$ fixed at 36~nm in all simulations. The results indicate that the thickness of the aluminium oxide layer that forms by natural oxidation under the gates $t_{\ce{Al2O3}}$ has a strong effect on the $J_{\rm direct}$ coupling. As $t_{\ce{Al2O3}}$ is increased from 1 to 5~nm, $J_{\rm direct}$ increases by almost three orders of magnitude when $d$ = 15~nm, compared to 1 order of magnitude when $d$ = 10~nm. We observed a similarly increasing trend of $J_{\rm direct}$ while increasing $t_{\ce{SiO2}}$ from 3--10 nm, as shown in Figure~\ref{fig3}(b). In this simulation, we have assumed $t_{\ce{Al2O3}}$ = 1~nm and $d = $15~nm. This phenomenon can be explained by assuming a constant lateral width of the aluminum gate, while varying the dielectric thickness. As the ratio of the dielectric thicknesses to the aluminum gate width increases, the electrostatic confinement of the dots reduces significantly, resulting in the electron wavefunctions expanding laterally and overlapping more. As $t_{\ce{SiO2}}$ is a highly controlled ultra-clean oxidation process, it suggests that we could vary the gate oxide thickness to obtain a specific coupling between the qubits. Alternatively, we could also use atomic layer deposited (ALD) oxide and vary the dielectric thickness. 

Figure~\ref{fig3}(c) shows that the $J_{\rm direct}$ coupling between qubits decreases as the oxidation thickness between gates $d$ increases, as expected. The large direct exchange coupling measured in the experimental spin funnels shown in Figure~\ref{fig2}(a) match the simulations for $d$ = 23.8~nm. Figure~\ref{fig3}(d) shows the simulated spin funnels, with a tunnel coupling of 957.7~MHz, very close to the experimental value $t_0$ = 900~MHz. 

%??Take note that the required $d$ =15??~nm is slightly higher than usual in order to get a close match to the experiment. Ideally, we expect $d$ < 10~nm as each aluminum gate does not oxidize more than 5~nm. Such discrepancy is due to the lack of information related to the measured device. The device specifications used in the simulation are based on our best understanding but they may vary from device to device. For instance, the dots size which have formed the qubits could be smaller than expected. It is not surprising as we have previously measured an 8~meV first orbital state energy for similar dots (cite Henry PRB), which has been fitted with < 20~nm dot diameter.

%Another potential cause could be charge traps induced in the low quality aluminum oxide giving similar effect to the tunnel barrier. A more comprehensive model is required to pinpoint the root cause of larger $d$ in order to obtain lower tunnel coupling between nearest neighbor qubits such as deploying random charge centers and modeling more realistic gate structures. This work is non-trivial and will be part of our future goal.

Here, the simulation parameters have been chosen based on our best understanding of the device's specifications but they may vary from device to device and affect the accuracy of the simulation results. For instance, it is hard to precisely determine the exact locations and sizes of the qubits. The separation between dot centres $x$ could be smaller or larger, as the quantum dots do not necessarily form underneath the center of the gates. The double dots could be confined nearer or further from each other, at the edges of the aluminum gates. Other inconsistencies could be caused by strain~\cite{thorbeck2015formation} induced at the oxide interface due to different thermal expansion of the aluminum gates and silicon when the device is cooled to 40~mK in the dilution refrigerator, and charged defects in proximity to the dots. Such strain can alter the conduction band and affect the tunnel coupling between the qubits. The charged defects close to the dots can also affect tunnel coupling~\cite{RR_PhysRevB.85.125423}. Finally, deviations from a perfectly flat (001) interface between Si and SiO$_2$ could lead to the quantum dot states presenting different valley phases, which would impact the exchange and potentially suppress it~\cite{li2010exchange}. Based on these simulated results, we note that although it is possible to design the optimum coupling required for multi-qubit operations, such efforts can be experimentally impractical due to uncertainties in the fabricated devices. As a solution, we could incorporate a $J$-coupling gate to tune the coupling between qubits when necessary, to maintain coherence and to assist in exchange~\cite{zajac2018resonantly,reed2016reduced}. 

Finally we note that, as expected, the exchange coupling depends exponentially on the gap between gates. This exponential decay of the direct exchange coupling with distance suggests that for second nearest neighbours $J_{\rm direct}$ may be comparable or even lesser than the mediated exchange coupling. While the direct mechanism for exchange is not ruled out, the observed coupling between Q1 and Q3 is most likely weaker than the coupling with the spin in Q2, which indicates that Q1 and Q3 interact through a three-body mediated coupling, as anticipated in previous theoretical studies~\cite{fei2012mediated,deng2020interplay}. 

%This is an important result for near term demonstrations of logical qubit error correction and prototypical quantum algorithms, in which a reduction in the necessary number of operations for coupling remote qubits may represent a more relaxed threshold for operation fidelities.

%According to ref~\citenum{thorbeck2015formation} and COMSOL simulation done in-house (results not shown), strain could change the conduction band energy by as much as 12~meV in the worst case scenario.

%Summary

%impractical to tune the J with different oxide thickness

In summary, we have demonstrated the addressability and control of three individual qubits in a linear array of silicon quantum dots. We have observed direct exchange coupling between nearest neighbour qubits, and much weaker, mediated super-exchange between next nearest neighbour qubits. Such coupling, if not engineered correctly, might lead to unwanted cross-coupling in multi-qubit devices. On the other hand, it may be leveraged for two qubit gates between remote qubits avoiding intermediate swap steps. NEMO simulations have uncovered the impact that the thickness of the gate dielectric has on the desired coupling, providing an engineering solution to adjust the coupling strengths. In particular, we find that increasing the dielectric thickness, separating the quantum dots from the electrostatic gates, increases the exchange coupling. This provides scope for fast qubit control executed at the charge symmetry point for high-fidelity operation.

\begin{acknowledgement}

We acknowledge support from the US Army Research Office (W911NF-17-1-0198), the Australian Research Council (FL190100167 and CE170100012), and the NSW Node of
the Australian National Fabrication Facility. The views and conclusions contained in this document are those of the authors and should not be interpreted as representing the official policies, either expressed or implied, of the Army Research Office or the U.S. Government. The U.S. Government is authorised to reproduce and distribute reprints for Government purposes notwithstanding any copyright notation herein. K. M. I. acknowledges support from a Grant-in-Aid for Scientific Research by MEXT, NanoQuine, FIRST, and the JSPS Core-to-Core Program. This research is part of the Blue Waters sustained-petascale computing project, which is supported by the National Science Foundation (awards OCI-0725070 and ACI-1238993) and the state of Illinois. Blue Waters is a joint effort of the University of Illinois at Urbana-Champaign and its National Center for Supercomputing Applications.

\end{acknowledgement}

\begin{suppinfo}

The gate oxide \ce{SiO2} is a thermally grown oxide, and is 5.9~nm thick, which is measured with an ellipsometer. The n$^{+}$ channels are diffused at high temperature using a phosphorus source. The aluminum gate stack is fabricated with three steps of electron beam lithography (EBL) with polymethyl methacrylate (PMMA) as mask and aluminum metal physical vapor deposition. In between each step, the aluminum gates are oxidized on a hotplate at 150~$^{\circ}$C for 10 minutes to form aluminum oxide as an insulating layer between overlapping gates. After all EBL steps, the sample is forming gas annealed with 5\% hydrogen and 95\% nitrogen mix gas at 3 L/min flow rate, at 400~$^{\circ}$C for 15 minutes to passivate charge traps at the oxide interface~\cite{do1988elimination}. All the above fabrication steps meet the industry standard for metal--oxide--semiconductor (MOS) processes.

Figure~\ref{figS1} depicts the Ramsey experiment on all three qubits to characterize their dephasing times, when loaded with one and three electrons. The dephasing times, $T_{2}^*$ are obtained by fitting the measurement to an exponentially decaying sinusoidal function, $f_{\uparrow} = a\exp^{(-\tau_d / T_{2}^*)}sin(\omega\tau_d+b)$, where $\tau_d$ is the delay time between two $X_{\pi/2}$ pulses in the Ramsey pulse sequences. For Q1, we obtained $T_{2}^*$ = 120~$\mu$s and 70~$\mu$s for one electron (red plot) and three electrons (blue plot), respectively. We highlight that the third electron has been loaded onto the upper valley state with the lower valley fully occupied by two electrons. We repeated the same Ramsey experiment for Q2 with one electron (pink plot) and Q3 with one (black plot) and three electrons (green plot). The $T_{2}^*$ for one electron loaded onto Q2 is 61~$\mu$s, whereas for Q3 are 30~$\mu$s and 55~$\mu$s with one and three electrons, respectively. We could not perform ESR on Q2 with three electrons as it is strongly coupled to one of the nearby dots in this configuration.

\begin{figure}
	\includegraphics[width=8.46cm]{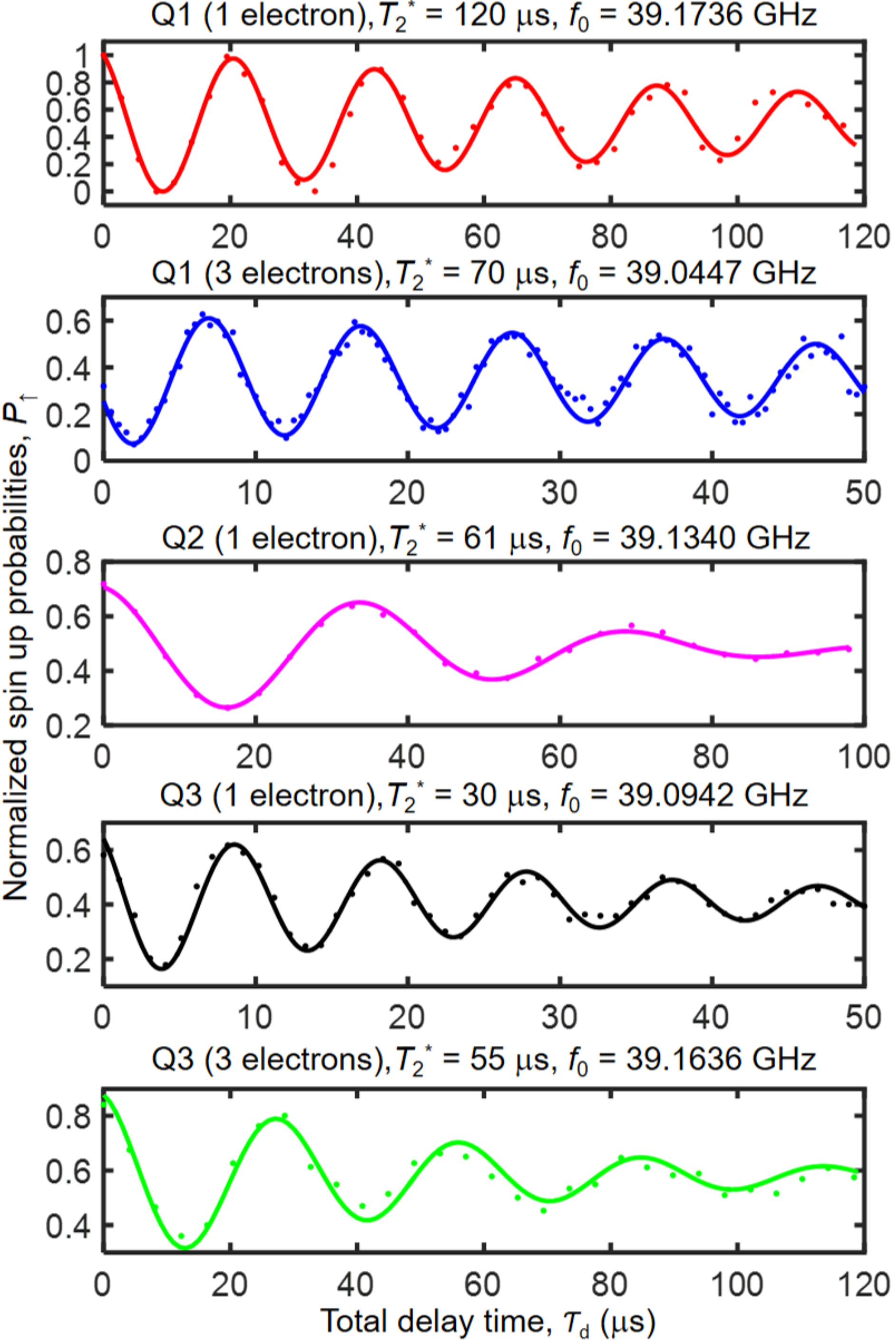}
	%  As well as the standard float types \texttt{table}\\
	%  and \texttt{figure}, the class also recognises\\
	%  \texttt{scheme}, \texttt{chart} and \texttt{graph}.
	\caption{Operation of the three-qubit system and their dephasing times, $T_{2}^*$ measured using Ramsey sequences. $T_{2}^*$ and ESR frequency, $f_0$ for each qubits Q1, Q2 and Q3, loaded with either one or three electrons as stated in their respective plots. We notice ESR at different frequencies for each qubit and when loaded with one or three electrons. The frequency $f_0$ for each of the three qubits varies by tens of MHz due to different gyromagnetic ratios caused by oxide interface roughness, and variabilities in the quantum dot geometries. These data indicate that we are able to control up to three qubits independently, in a linear array.
	}
	\label{figS1}
\end{figure}

From the measurement, each qubit has a different resonance frequency indicating the ability to address the qubit that one desires to manipulate. In the one-electron qubit system, we obtained $f_{0}$ = 39.1736~GHz, 39.1340~GHz and 39.0942~GHz for Q1, Q2 and Q3, respectively. As each qubit has its associated top-gate, individual Stark tuning is possible to create individual qubit addressability. The apparent Stark shift depends on the quantum dot and the electron number. We notice for the three-electron qubit, $f_{0}$ = 39.0447~GHz and 39.1636~GHz for Q1 and Q3. Interestingly, we observe that the resonance frequency is higher for Q1 with one-electron but higher for Q3 with three-electron loaded. A possible explanation is the presence of steps at the oxide interface, for example due to the finite miscut in the silicon wafer~\cite{PhysRevB.54.R2304,PhysRevLett.64.2406} and randomness in the $\ce{SiO2}$ thermal oxidation growth process. Such a step alters the valley composition and thereby strongly affects the impact of spin-orbit coupling and the renormalization of the $g$-factor~\cite{WisterPRB2017}. Experimentally we find that for Q1 the one-electron qubit performance is better, but for Q3 the three-electron qubit performance is better (see coherence times estimated from the Ramsey measurement). These results are consistent with different valley compositions for the two different dots, resulting in a different spin flip competition between the intra-valley and inter-valley spin-orbit coupling~\cite{Veldhorst_SOC_PRB2014}. 

The exchange coupling $J$ simulation framework consists of an FEM-based electrostatics solver, a tight-binding Hamiltonian solver, and a full configuration interaction solver. The factors affecting the electrostatic environment in the dots are device geometry, gate voltages, background impurities such as interface charge and doping, and the charges occupying the dots themselves. The electrostatic potential in the dots is evaluated by self-consistently solving the semiconductor Poisson equation with the effective-mass Schr\"odinger equation to account for quantum confinement effects, on a finite-element grid~\cite{sahasrabudhe2018optimization}.

The electrons occupying the dots are in the X-valleys of the silicon band-structure. The electric fields in the dots break the 6-fold valley degeneracy. Crystal symmetry breaking due to the interface, and roughness in the silicon oxide interface leads to a enhanced spin-orbit coupling. Consideration of these atomistic features is required for accurately obtaining the energy spectrum and wavefunctions. This is done by solving an atomistic tight-binding Hamiltonian with the electrostatic potential calculated in the previous step.

The few-electron wavefunctions are then obtained using the full configuration interaction method on the tight-binding basis. The few-electron Hamiltonian includes higher orbital and higher valley components of the tight-binding spectrum as a result, and captures exchange and correlation effects in detail.~\cite{PhysRevB.97.195301}.

\end{suppinfo}

\bibliography{manuscript}

\end{document}